\documentclass[superscriptaddress,preprintnumbers,keywords]{revtex4}
\usepackage{epsfig,dcolumn}
\usepackage{graphicx}
\usepackage{color}

\usepackage{amsmath}
\usepackage{amsfonts}
\usepackage{amssymb}
\usepackage{graphicx}

\newcommand{\be}{\begin{equation}}
\newcommand{\ee}{\end{equation}}

\newcommand{\eea}{\end{eqnarray}}

\def\lsim{\mathrel{\rlap{\lower4pt\hbox{\hskip1pt$\sim$}}
    \raise1pt\hbox{$<$}}}
\def\gsim{\mathrel{\rlap{\lower4pt\hbox{\hskip1pt$\sim$}}
    \raise1pt\hbox{$>$}}}

\usepackage{bm} 

\begin{document}

\title{Dalitz plot distributions in presence of triangle singularities} 

\author{Adam P. Szczepaniak} 
 \affiliation{ 
  Department of Physics, Indiana University, Bloomington, IN 47405, USA } 
\affiliation{ Theory Center, Thomas Jefferson National Accelerator Facility, \\
12000 Jefferson Avenue, Newport News, Virginia 23606, USA} 
  \affiliation{ 
 Center for Exploration of Energy and Matter, Indiana University, Bloomington, IN 47403, USA}

\preprint{JLAB-THY-16-1}


\begin{abstract}
We discuss properties of three-particle Dalitz distributions in coupled channel systems in presence of triangle singularities.  The single channel case was discussed long ago  ~\cite{schmid} where it was found that as a consequence of unitarity,  effects of a triangle singularity seen in the Dalitz plot are not seen 
 in Dalitz plot projections. 
  In the coupled channel case we find the same is true 
 for the sum of intensities of all interacting channels. Unlike the single channel case, however, triangle singularities do remain visible in Dalitz plot projections of individual channels. 
  \end{abstract}

\maketitle

\section{Introduction}
\label{sec:introduction}
 
\begin{figure}
\centering
\rotatebox{0}{\scalebox{0.25}[0.25]{\includegraphics{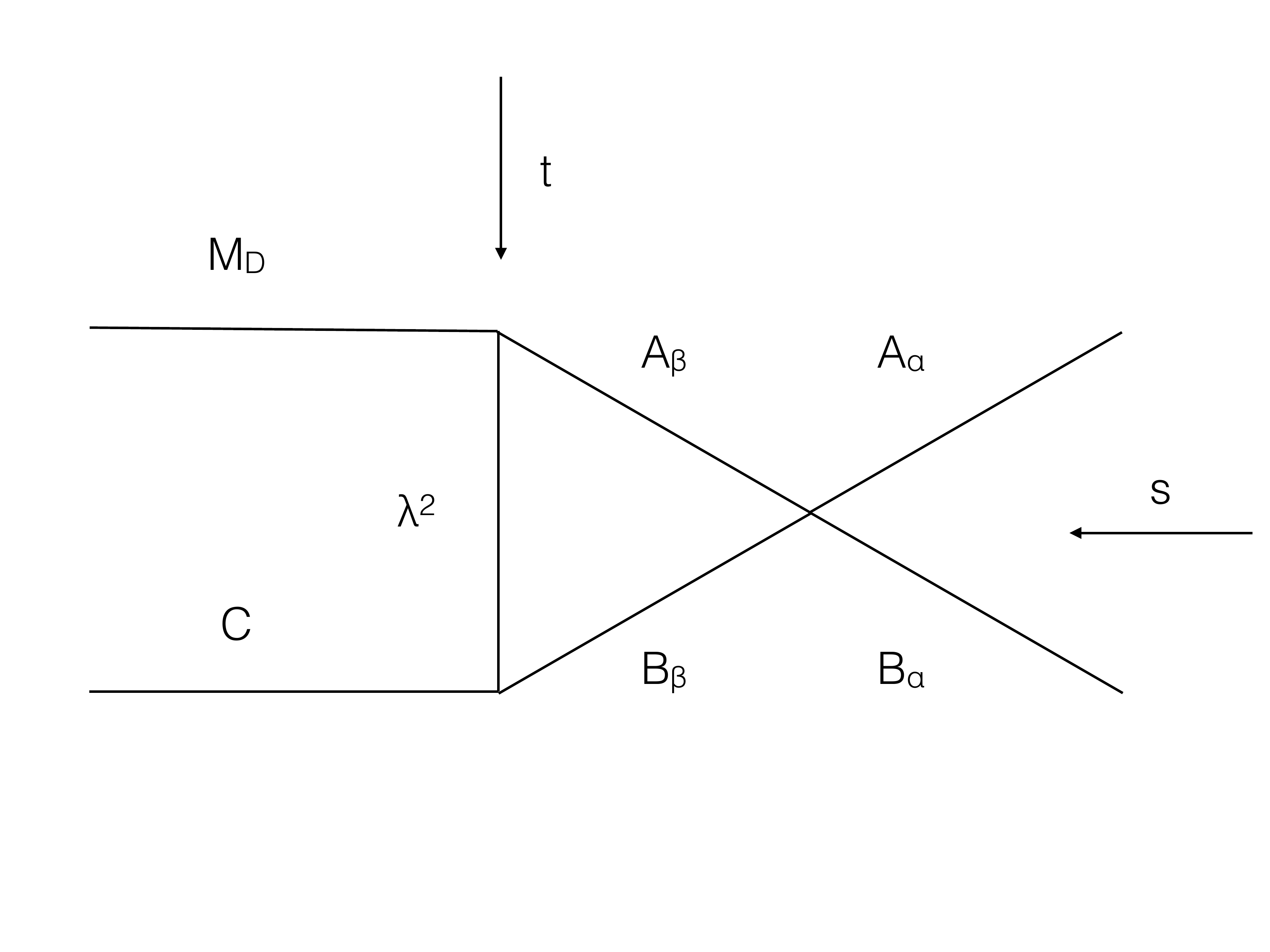}}}
\caption{ Triangle  diagram representing the process  $D \to A_\alpha B_\alpha C$ with a $t$-channel exchange of a pole at $t=\lambda^2 - i\epsilon$ with couple channel interactions in the $s$-channel. } 
\label{fig:1}
\end{figure}

 Under specific kinematic conditions~\cite{CN}, triangle diagrams~\cite{Aitchison:2015jxa} have singularities that can mimic resonance poles.  For this reason partial wave peaks at energies that do not 
  match the known hadron spectrum   {\it e.g.}  as  expected from the quark model, have occasionally been attributed to such effects. Most recently, for example, triangle singularities have been discussed  in the context of the XYZ  quarkonium peaks~\cite{Wang:2013cya,Chen:2011pv,Chen:2011xk,Chen:2013coa,Guo:2014qra,Pakhlov:2014qva,Guo:2014iya,Szczepaniak:2015eza},  the peak in the $J^{PC}=1^{++}[\rho\pi]$ partial wave~\cite{Ketzer:2015tqa},  {\it i.e.} the $a_1(1420)$  seen in the COMPASS data on pion diffractive dissociation~\cite{Adolph:2015pws}, or  the pentaquark signal~\cite{Guo:2015umn, Mikhasenko:2015vca} reported by the LHCb collaboration~\cite{Aaij:2015tga}.  Triangle singularities have a simple interpretation when the underlying amplitude is expressed as a dispersive integral. In Fig.~\ref{fig:1} we show a triangle diagram describing decay of a quasi-stable particle $D$ of mass $M_D$ to three stable particles, $A_\alpha$, $B_\alpha$, $C$ through  coupling to a pair of particles $A_\beta, B_\beta$.  In the following, for simplicity, we ignore all particle spins and  consider a case of two coupled two-body channels, ( $\alpha,\beta = 1,2$).  The triangle diagram can be expressed through a dispersive integral in which the  on-shell amplitude describing  $t$-channel exchange of a particle of mass $\lambda$ is projected onto the $s$-channel partial wave and unitarized. The projected amplitude  (in the next section denoted by $b_{l,\alpha}(s)$) has two of its four branch points moving  as a function of $\lambda$~\cite{Szczepaniak:2015eza}. For a range of (real) $\lambda^2$, determined by the Coleman-Norton condition ~\cite{CN}, one of these branch points, $s_T$ is located infinitesimally below the real $s$-axis and above the $s$-channel threshold, $s_\beta$. This leads to a logarithmic branch point in the dispersive integral located on the second sheet just below the physical region (the physical region is defined as $s + i\epsilon$). 
  The triangle singularity is constrained by the two-body unitarity.  The Coleman-Norton condition requires 
   $\lambda \ge B + C$. Taking into account $t$-channel unitarity this 
   implies that only resonances (and not stable particles)  are involved. 
 Due to the finite resonance width the singular point $s=s_T$ is shifted away from the physical region down the $s$-channel unitary cut and onto the second sheet.\footnote{ If the singularity was located on the physical axis it would violate the $s$-channel unitarity.}  
   The  analysis is similar to that of the standard  Muskhelishvili-Omnes problem ~\cite{Omnes:1958hv,Mandelstam:1960zz,Pham:1976yi}   with the only difference being that in the case considered here  
     the left hand cut is actually located in the complex $s$-plane and for narrow $t$-channel resonances may be close to the physical region, {\it i.e.} near the right hand cut. In other applications of 
 triangle diagrams, however, two-body unitarity is not sufficient. For example in the analysis of the $a_1(1420)$~\cite{Ketzer:2015tqa} the $t$-channel exchange of a stable kaon connects the  $f_0(980) \pi$
and   $K^* \bar K$,      {\it aka} $ K \bar K \pi$  three-particle states. 
 In this cases it is necessary to invoke three-body unitary to constrain the triangle amplitude. 
   
  In the following we give a detailed discussion of the coupled  Muskhelishvili-Omnes (MO) problem in presence of triangle  singularities. In particular we determine what type of structures are 
 to be expected in the Dalitz plot distributions. 
   The single channel case was discussed in ~\cite{schmid} and revisited in  ~\cite{AA}.  The reason why generalization to coupled channels is 
    of interest  is because, for example, the XYZ phenomena tend to occur in vicinity of several  open quasi-two-body channels.

\section{Combining $s$, $t$, and $u$, channel isobars} 
\label{sec:model} 

 We are interested in amplitudes describing a decay of a  quasi-stable particle $D$ with mass $M_D$ to 
  two channels, $\alpha=1,2$ of three distinguishable particles  $A_\alpha,B_\alpha,C$.  
     The decay amplitude $A_\alpha(s,t,u)$, depends on 
  the three Mandelstam invariants, which we define as $s = (p_A + p_B)^2$, $t = (p_B + p_C)^2$ and $u=(p_A + p_C)^2$ and are kinematically constrained by $s + t + u = \sum_i m_i^2$. 
 Analyticity of the  $S$-matrix implies that, besides the decay channel, the same amplitude describes each of the three  two-to-two scattering processes, {\it i.e} the $s$-channel reaction $D + \bar C \to A + B$, (bar denotes an antiparticle) as well as the $t$ and $u$ channel scattering.  Therefore, the amplitude in the physical domain of the  decay process can be  obtained by analytical continuation of the amplitude from, say the $s$-channel scattering physical region. 
    Partial wave expansion in the $s$-channel, 
  \begin{equation} 
  A_\alpha(s,t,u) = \frac{1}{4\pi} \sum_l (2l+1) f_{l,\alpha}(s) P_l(z_s) \label{spw} 
  \end{equation} 
  with $z_s$ being cosine of scattering angle, converges in the $s$-channel physical region and in the decay region ($|z_s| < 1$). In the $s$-channel physical region,  complexity of the partial waves, $f_{l,\alpha}(s)$ is determined by  $s$-channel singularities. 
      In the decay channel, however, in addition to the $s$-channel,  $t$ and $u$ channel singularities are  also physical and contribute to the complexity of the $s$-channel partial waves. It follows that
       in order to use Eq.~(\ref{spw}) in the kinematical region of the decay process, the sum on {\it r.h.s}
        has to be analytically continued. 
 Therefore a finite set of $s$-channel partial waves cannot reproduce $t$ or $u$-channel singularities, {\it e.g.} a  resonance that appears inside the Dalitz plot. In the isobar model, in which a finite number of $s$-channel partial waves is considered,  the omitted infinite number of waves is replaced by a finite number of $t$ ad $u$ waves.  The  amplitude has a mixed form that includes partial waves (isobars) in the three channels simultaneously, 
 
 \begin{equation}
 A_\alpha(s,t,u) = A^{(s)}(s) + A^{(t)}(t) + A^{(u)}(u), \;  
 A^{(x)}(x) =  \frac{1}{4\pi} \sum_{l=0}^{L_{max}} (2l+1) a^{(x)}_{l,\alpha}(x) P_l(z_x), \; x=s,t,u.  
  \label{im} 
 \end{equation} 
 We refer to the amplitudes $a^{(x)}_{l,\alpha}(x)$ as the isobaric amplitudes in the $x$'th channel. 
  The isobaric amplitudes, say in the $s$-channel, $a^{(s)}_l(s)$ contain the $s$-channel unitary cut and may also contain left hand cuts. To avoid double counting, however,  the latter should not overlap  
 with the cuts that originate from projections onto the $s$-channel partial waves of the $t$ and $u$-channel isobars. 
 In the following we ignore any remaining, distant left hand cuts of the isobaric amplitudes. 
 In a  Dalitz plot analysis, the isobaric amplitudes are typically parametrized using  
  energy dependent Breit-Wigner formulae but this can be easily generalized~\cite{as}. 
 
 We examine  implications of a triangle singularity present in the $t$-channel in one of the two channels, 
  {\it e.g.} in $D + \bar A_1 \to B_1 + C$ and ignore the $u$-channel exchange contributions, {\it e.g.} set $A^{(u)} = 0$.  For simplicity,  we also assume that only $S$-wave ($l=0$) interactions between pars $A_\alpha,B_\alpha$ are strong and are given by a $2\times 2$ set of unitary, $S$-wave amplitudes 
   $t_{0,\alpha\beta}(s)$, satisfying, 
 
 \begin{equation} 
\Delta t_{0,\alpha\beta}(s) = Im t_{0,\alpha\beta} =  \sum_{\gamma=1,2} t^*_{0,\alpha,\gamma}(s) \rho_\gamma(s) t_{0,\gamma\beta}(s).
\end{equation} 
 Here $\Delta$ denotes the right hand cut discontinuity, and $\rho_\alpha(s)$ is the channel phase space 
$ \rho_\alpha(s) = \lambda(s,m^2_{A_\alpha},m^2_{B_\alpha})/2\sqrt{s}$ with $\lambda$ being the triangle function.  Projecting the {\it r.h.s} of Eq.~(\ref{im}) onto the $s$-channel gives the partial wave expansion of the model,  
 \begin{equation} 
 f_{l,\alpha}(s)  = a_{l,\alpha}(s) +  b_{l,\alpha}(s), 
\end{equation} 
with $a_{l,\alpha}(s) =  a^{(s)}_{l,\alpha}(s)$ and nonzero only for $l=0$. For all $l$'s, 
\begin{equation}
b_{l,\alpha}(s)  = \frac{1}{2} \int_{-1}^1  dz_s P_l(z_s) \sum_{l'=0}^{L_{max}} (2l' + 1) a^{(t)}_{l',\alpha}(t+i\epsilon) P_{l'}(z_t).  
 \end{equation} 
 Under the integral, $t$ and $z_t$, the cosine of the $t$-channel scattering angle, are to be considered as  functions of $s$ and $z_s$.  The amplitude  $b_{l,\alpha}(s)$ is the $s$-channel projection of  $t$ channel exchanges and has complex singularities in the $s$-plane.   The location of these singularities is determined by unitarity in the $t$-channel.  Unitarity leads to an amplitude  that is analytical in the $t$-channel physical region {\it i.e.} for $t$ infinitesimally above the real axis. 
 Note that there is no need to make $M_D$ complex. Singularities of the amplitude in the $M_D$ variable
   are controlled by the three-particle unitarity. When three-body unitarity is involved the diagram considered here plays a role of a vertex function which should have no unitary cuts in $M_D$.

\begin{figure}
\centering
\rotatebox{0}{\scalebox{0.25}[0.25]{\includegraphics{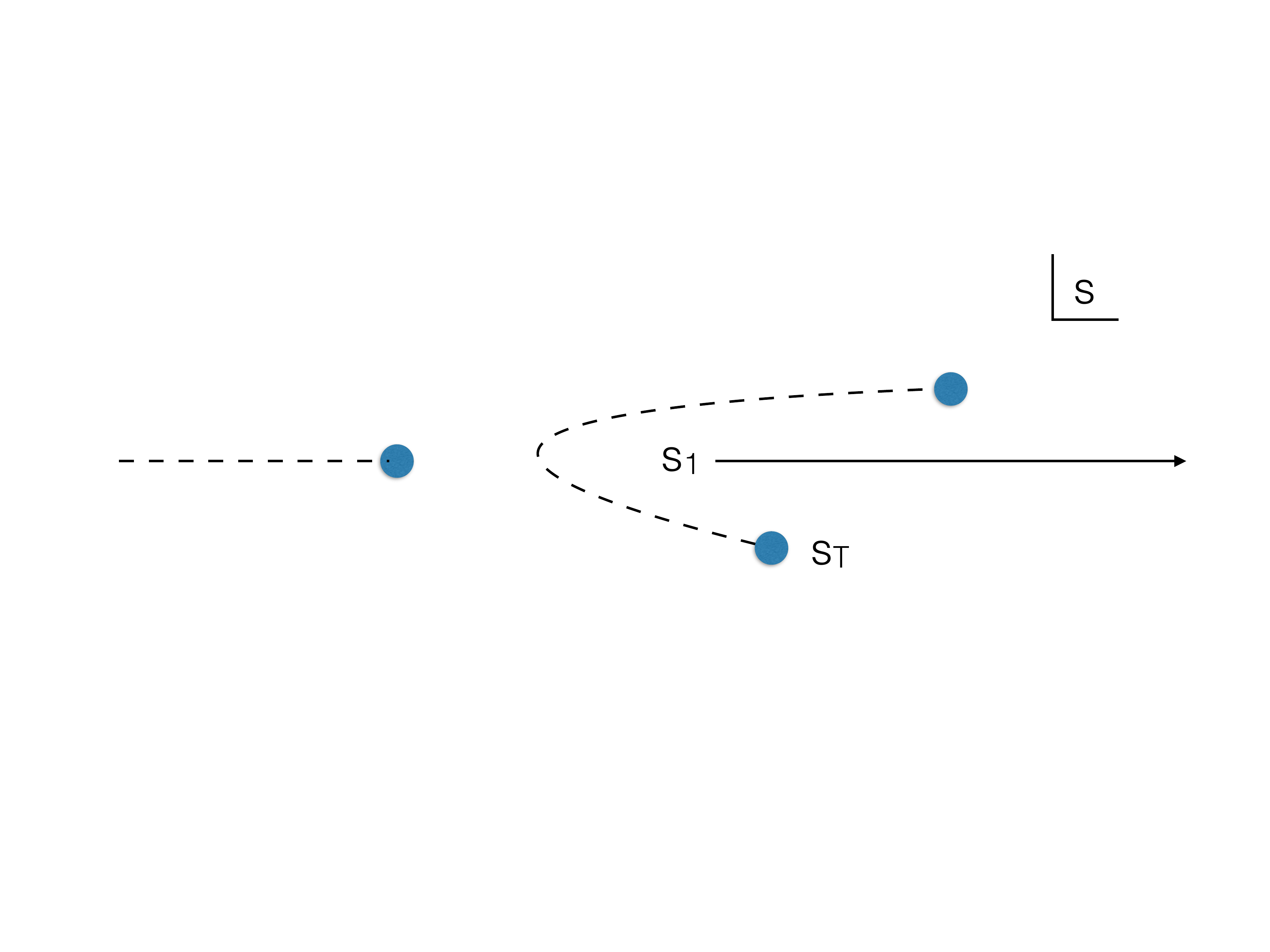}}}
\caption{ Location of cuts (dashed lines) of the amplitude $b_{0,1}(s)$ in the complex $s$  plane. The triangle singularity is due to the $s_T$ branch point located below the real $s$-axis and to the right of the channel-1 threshold, $s_1$.} 
\label{fig:2}
\end{figure}

  
  Unitarity in the $s$-channel determines discontinuity of the $f_{l,\alpha}(s)$, partial wave across 
   the right hand cut. With the assumption, that $A_\alpha$ and $B_\alpha$ interact strongly in 
   the $S$-wave only we find, 

\begin{equation}
\Delta f_{0,\alpha}(s)   = \Delta a_0(s) = \sum_{\beta=1,2} t^*_{\alpha,\beta} \rho_\beta(s) f_{0,\beta}(s), \;
\Delta f_{l,\alpha}(s) = 0, \mbox{ for } l > 0.
\label{ua}
\end{equation} 
 The reason why it is $\Delta f$ and not $Im f$ appears on the {\it l.h.s} of the unitary equation is  
  the decay kinematics. As discussed below Eq.~(\ref{spw}), cross channel exchanges are 
  physical in the $s$-channel and lead to additional (beyond the one determined by $s$-channel unitarity)  complexity of the $s$-channel partial waves.  As a function of $s$, the projected amplitudes, 
      $b_{l,\alpha}(s)$  have the left hand cut but do not have the right hand $s$-channel unitary  cut. 
       In particular, in presence of triangle singularities, when the Coleman-Norton conditions are met, \cite{CN}, a portion of the left hand cut of $b_{l,\alpha}(s)$  surrounds the $s$-channel threshold branch point as illustrated   in Fig.~\ref{fig:2}.

 From  Eq.~(\ref{ua}) it follows that, for $l > 0$, modulo distant left hand cuts or subtractions, 
 \begin{equation}
 f_{l,\alpha}(s) = b_{l,\alpha}(s)
 \end{equation} 
 while for the  $S$-wave, 
\begin{equation} 
  f_{0,\alpha}(s) = \sum_{\beta = 1,2} t_{0,\alpha,\beta}(s)  G_{0,\beta}(s) \label{sf} 
  \end{equation} 
 with the function $G_{0,\beta}(s)$ given by 
  \begin{equation}  
  G_{0,\alpha}(s) =  \frac{1}{\pi} \int_{s_{\alpha}} ds' \rho_\alpha(s') \frac{  b_{0,\alpha}(s') - b_{0,\alpha}(s)}{s' - s} \label{sf1} 
\end{equation} 
 The function $G_{0,\alpha}(s)$ is free from the $s$-channel unitary cut. The above form was derived assuming the isobaric amplitudes have no left hand cuts. It is, however,  straightforward to generalize Eq.~(\ref{sf1})  to include such cuts.  Even though $G_{0,\alpha}(s)$ is continuous across the unitary cut, as explained  below Eq.~(\ref{ua}), even in the single channel case, it is complex on the real axis all the way down to threshold $s_\alpha$. In the decay kinematics the Watson theorem $Im f_0(s)  =  Im t(s)$ does not apply. Instead  it is replaced by the relation for the discontinuity given in Eq.~(\ref{ua})~\cite{Guo:2014vya}.


\section{ Implication for Dalitz plot distributions} 

Finally we are in the position to examine the consequences of the $s$-channel unitarity constraint, {\it cf.}  Eq.~(\ref{sf}) on the distribution of events in the Dalitz plot for both decay 
 channels.  We rewrite the $S$-wave partial wave amplitude as, 
 \begin{equation} 
   f_{0,\alpha}(s) = \sum_{\beta= 1,2} t_{0,\alpha\beta}(s) \left[ \sum_{\gamma=1,2} [t^{-1}_0]_{\beta \gamma}(s) [b_{0,\gamma}(s) + g_{0,\gamma}(s)]  
  +  \frac{1}{\pi} \int_{s_{\beta}} ds' \rho_\beta(s') \frac{  [b_{0,\beta}(s')  + g_{0,\beta}(s')] }{s' - s}  \right]. 
  \end{equation}
 The functions $g_{0,\alpha(s)}$ are determined by the left hand discontinuities of $t_{0}(s)$ imposing 
  the condition that isobaric amplitude have no  left hand cuts~\cite{Guo:2014vya}. 
 Using the unitarity relation for $t_0$ we obtain 
\begin{equation}
  f_{0,\alpha}(s) = \sum_{\beta = 1,2} t_{0,\alpha\beta}(s) \left[ \sum_{\gamma=1,2} [t^{-1*}_0]_{\beta\gamma}(s) [b_\gamma(s) + g_\gamma(s)]  
   - 2 i \rho_\beta(s) [ b_{0,\beta}(s)  + g_{0,\beta}(s)]+ \frac{1}{\pi} \int_{s_{\beta}} ds' \rho_\beta(s') \frac{  [b_{0,\beta}(s')  + g_{0,\beta}(s')] }{s' - s}  \right].  
  \end{equation}
 We assume that the second sheet logarithmic branch point, $s_T$ and the thresholds $s_\alpha$ are 
   the only relevant singularities in the vicinity of the physical region. The triangle singularity occurs under a very constrained kinematics, thus it is safe to assume that it occurs in one channel, {\it e.g.}  $\alpha=1$ only.   On the sheet connected to the physical region one finds 
\begin{equation} 
 \lim_{s \to s_T} \frac{1}{\pi} \int_{s_{1}} ds' \rho_1(s') \frac{  [b_{0,1}(s')  + g_{0,1}(s')] }{s' - s} 
    =  2 i \rho_1(s_T) \lim_{s \to s_T} b_{0,1}(s) + \cdots 
\end{equation} 
where the ellipsis denotes  terms that are finite in the limit $s \to s_T$ ($s_T$ is in the complex plane). 
  In the following we ignore such terms. 
In terms of the $S$-matrix, whose $l=0$-partial wave,   $2\times 2$ channel  matrix elements are given in terms of the $t_0$ matrix elements by, 
\begin{equation} 
S_{0,\alpha\beta}(s)  = \delta_{\alpha\beta} + 2i \sqrt{\rho_\alpha(s)} t_{0,\alpha\beta}(s) \sqrt{\rho_\beta(s)} 
\end{equation} 
one easily finds that terms that are singular at $s=s_T$,  in the physical region (real-$s$) give, 
\begin{equation}  
 f_{0,1}(s) =   S_{0,11}(s) b_1(s), \; 
 f_{0,2}(s) = \frac{\sqrt{\rho_1(s)}}{\sqrt{\rho_2(s)}} S_{0,21}(s) b_1(s) 
\end{equation} 
while the higher partial waves are given by, 
\begin{equation} 
f_{l,1}(s) = b_{l,1}(s), \; f_{l,2}(s) = b_{l,2}(s)  = \cdots .
\end{equation} 
{\it i.e.}  $f_{l,2}$ having no near-by triangle singularities.  
Thus  assuming all $s$-channel interactions are negligible except in the $S$-wave 
 and that the $s_T$ singularly appears in  channel $1$ only, we find ({\it cf.} Eq.~(\ref{im})) 
\begin{eqnarray} 
& & A_1(s,t,u) = \frac{1}{4\pi} [S_{0,11}(s) - 1] b_{0,1}(s)   + A^t_1(t)  \nonumber \\
& & A_2(s,t,u) = \frac{1}{4\pi}  \sqrt{\frac{\rho_1(s)}{\rho_2(s)}} S_{0,21}(s) b_{0,1}(s)  + A^t_2(t) \label{As} 
\end{eqnarray} 

In the channel $\alpha$ the Dalitz plot intensity distribution is proportional to the magnitude squared  of the decay amplitude, 
\begin{equation} 
I_\alpha(s,t,u) =  |A_\alpha(s,t,u)|^2.  
\end{equation} 
If follows that in both Dalitz plots there will be an $s$-channel band that originates from $s$-dependent, 
  first term on the  {\it r.h.s} of Eq.~(\ref{As}).  As for the $s$-channel projection of the Dalitz plot, which is proportional to 

 \begin{equation}
 I^{(s)}_\alpha(s)  =   \rho_\alpha(s) \int_{-1}^1  dz_s  I(s,t,u) = \frac{\rho_\alpha(s)}{8\pi^2}  \sum_l (2l+1)|f_{l,\alpha}(s)|^2 
 \end{equation} 
we find 
\begin{eqnarray} 
& & I_1^{(s)}(s) \propto \left[  |S_{11}(s)|^2 - 1 \right]  \rho_1(s) |b_{0,1}(s)|^2 + \cdots  \nonumber \\
& & I_2^{(s)}(s) \propto  |S_{21}(s)|^2  \rho_1(s) |b_{0,1}(s)|^2  + \cdots  \label{pro} 
 \end{eqnarray}
where, as before, ellipses indicate contributions regular in the limit $s \to s_T$.  This is our main result. 
 When reduced to the single channel case, by setting $|S_{11}| = 1$ and $S_{21} = 0$, it reproduces the result of ~\cite{schmid}. Namely, the absence of an enhancement in the $s$-channel Dalitz plot projection 
 due to a triangle singularity, even though (for $Im S_0 \ne 0$)   it  can produce  a visible band in the Dalitz plot. 

      \section{Summary} 
     It follows from Eq.~(\ref{pro})  that in the couple channel case, the result of ~\cite{schmid} generalizes. One finds that the net sum of events in the $s$-channel Dalitz plot projections of the two coupled channels, {\it i.e.} $\sum_{\alpha} I^{(s)}_\alpha(s)$  does not display variation as a function of $s$ due to the triangle singularity.    On the other hand, contrary to what happens in the single channel case, 
      Eq.~(\ref{pro}) predicts that the effect of a triangle singularity should be visible in projections of Dalitz distributions in individual channels.   Specifically for $s$ near the band, (as  determined by the location of $s_T,$)  with the singularity appearing in channel $1$,  one expects to see enhancement  
       in the Dalitz projection of  channel-$2$ and reduction in events in the projection of  channel $1$. 
    The former was  observed,  for example,  in the analysis of the decay  $Y(4260)  \to \pi^+\pi^- J/\psi$~\cite{Szczepaniak:2015eza}.  The $Z_c(3900)$ peak was attributed to  the triangle singularity 
 emerging from the $t$-channel exchange of $D_0(2400)$ coupled to the $D\bar D^*$ (channel $1$)  
 re-scattering  $\pi J/\psi$ (channel $2$). According to Eq.~(\ref{pro}) the crucial test of this
    hypothesis  is to verify that  the $Y(4260) \to D \bar D^* \pi + {\it c.c.}$ Dalitz plot projection is depleted in $s$ in the vicinity of the $Z_c(3900)$.

\section*{Acknowledgments}
I would like to thank   Ian Aitchison for the discussions during the 
International Summer School on Data Science for Scattering Reactions at Indiana University in Bloomington, IN and his suggestion to revisit this problem.  I would like to thank Andrew Jackura for his comments on the manuscript.  The School was supported by the NSF under contract PHY-1513524. This  work is supported by the U.S. Department of Energy, Office of Science, Office of Nuclear Physics under contract DE-AC05-06OR23177. It is  also supported in part by the U.S. Department of Energy under Grant No. DE-FG0287ER40365.

\end{document}